\documentclass[twocolumn,superscriptaddress,nofootinbib,aps,prl,floatfix,preprintnumbers,amsmath,amssymb,groupedaddress]{revtex4-1}

\pdfoutput=1 

\usepackage{amsmath}
\usepackage{dsfont}
\usepackage{epsfig}
\usepackage{slashed}
\usepackage{bbold}
\usepackage{psfrag}
\usepackage{color}
\PassOptionsToPackage{caption=false}{subfig}
\usepackage{subfig}
\usepackage{multirow}
\usepackage{booktabs}
\usepackage{hyperref}
\usepackage{pstricks}
\usepackage{feynmp}
\usepackage{soul}

\usepackage{color}
\usepackage[normalem]{ulem} 
\usepackage[utf8]{inputenc}

\newcommand{\beq}{\begin{equation}}
\newcommand{\eeq}{\end{equation}}
\newcommand{\bea}{\begin{eqnarray}}
\newcommand{\eea}{\end{eqnarray}}

\newcommand{\DE}{\Delta E_B}
\newcommand{\sigmap}{\bar{\sigma}_n}

\newcommand{\FdisE}{\ifmmode {\lvert F_{dis}(E_r,E_{int}) \rvert}^2 \else ${\lvert F_{dis}(E_r,E_{int}) \rvert}^2$\fi}
\newcommand{\Fmol}{\ifmmode {\lvert F_{mol}(\mathbf{q,\tilde{\mathbf{q}}}) \rvert}^2 \else ${\lvert F_{mol}(\mathbf{q},\tilde{\mathbf{q}}) \rvert}^2$\fi}
\newcommand{\FDM}{F_{\rm DM}(\mathbf{q})}
\newcommand{\FDMq}{F_{\rm DM}(q)}
\newcommand{\FDMsq}{|F_{\rm DM}(\mathbf{q})|^{2}}

\newcommand{\Mfi}{\ifmmode {\lvert \mathcal{M}_{2-2} \rvert}^2 \else ${\lvert \mathcal{M}_{2-2} \rvert}^2$\fi}

\newcommand{\units}[1]{\mathrm{\; #1}}

\renewcommand{\subsubsection}[1]{\addtocounter{subsubsection}{1}
\par\nobreak
\medskip
\nobreak
\noindent{\it \thesubsubsection.  #1 }
\par\nobreak\medskip\nobreak}

\def\lpar#1#2#3#4{\rlap{\raise#3\hbox{$\hskip#4#1\left\{\mbox{\phantom{\rule[0mm]{0mm}{#2}}}\right.$}}}
\def\rpar#1#2#3#4{\rlap{\raise#3\hbox{$\hskip#4\left\}#1\mbox{\phantom{\rule[0mm]{0mm}{#2}}}\right.$}}}

\begin{document}

\title{Direct Detection of Light Dark Matter and Solar Neutrinos via Color Center Production in Crystals}

\author{Ranny Budnik}
\affiliation{Department of Particle Physics and Astrophysics, Weizmann Institute of Science, Rehovot, Israel}
\email{ran.budnik@weizmann.ac.il}

\author{Ori Chesnovsky}
\affiliation{Raymond and Beverly Sackler School of Chemistry, Tel-Aviv University, Tel-Aviv, Israel}
\email{orich@post.tau.ac.il}

\author{Oren Slone}
\affiliation{Raymond and Beverly Sackler School of Physics and Astronomy, Tel-Aviv University, Tel-Aviv, Israel}
\email{shtangas@gmail.com}

\author{Tomer Volansky}
\affiliation{Raymond and Beverly Sackler School of Physics and Astronomy, Tel-Aviv University, Tel-Aviv, Israel}
\email{tomerv@post.tau.ac.il}

\begin{abstract}
We propose a new low-threshold direct-detection concept for dark matter and for coherent nuclear scattering of solar neutrinos, based on the dissociation of 
atoms
and subsequent creation of color center type defects within a lattice. The novelty in our approach lies in its ability to detect single defects in a macroscopic bulk of material. This class of experiments features ultra-low energy thresholds which allows for the probing of dark matter as light as $\mathcal{O}(10)$ MeV through nuclear scattering. Another feature of defect creation in crystals is directional information, which presents as a spectacular signal and a handle on background reduction in the form of daily modulation of the interaction rate. We  discuss the envisioned setup and detection technique, as well as background reduction.  We further calculate the expected rates for dark matter and solar neutrinos in two example crystals for which available data exists,  demonstrating the prospective sensitivity of such experiments. \end{abstract}

\maketitle

\section{Introduction} 
\label{sec:Intro}

More than 80\% of the matter in our universe is yet to be discovered.
This astonishing fact has been established with overwhelming evidence by measurements ranging from sub-galactic to cosmological scale. Yet so far, this so called Dark Matter (DM) has been manifested via gravitational interactions only, and its particle nature remains unknown.

For over three decades there has been an extensive effort to search for DM directly with underground detectors, indirectly with the use of satellites and earth-based telescopes, and at colliders such as the Large Hadron Collider (LHC). To date there is no unambiguous, non-gravitational, experimental evidence for DM. Most of the theoretical and experimental effort, however, has been focused on a specific DM paradigm, the Weakly Interacting Massive Particle (for a review, see e.g.~\cite{Bertone:2004pz}). While appealing, the failure to discover the WIMP suggests that an alternative DM scenario may be at play. The last half a decade has seen significant theoretical progress in this direction, with a particularly motivated candidate being Light Dark Matter (LDM) in the MeV to GeV mass range~\cite{Boehm:2003hm, Boehm:2003ha, An:2014twa,An:2013yua, Feng:2008ya, Hooper:2008im, DAgnolo:2015ujb, Kusenko:2009up, Kaplan:2009ag, Essig:2010ye, Choi:2011yf, Falkowski:2011xh, Lin:2011gj, Hochberg:2014dra, Hochberg:2014kqa, Boddy:2014yra,Rajagopal:1990yx, Covi:1999ty}. Despite a number of proposals for experimental setups for LDM direct detection over the last years~\cite{Essig:2012yx,Essig:2015cda,Graham:2012su,Lee:2015qva,Hochberg:2015pha,Hochberg:2015fth,Hochberg:2016ajh,Hochberg:2016ntt,Hochberg:2016sqx,Schutz:2016tid,Kouvaris:2016afs,Bloch:2016sjj,Bunting:2017net,Derenzo:2016fse,Knapen:2016cue,McCabe:2017rln}
, currently almost no experimental opportunities exist to search for such LDM. In this study, we propose  a novel experimental technique to directly search for LDM with masses below that of the proton. Our suggested setup will allow to significantly expand the experimental effort  beyond current capabilities.

DM direct detection experiments typically search for the small recoiling energy imprinted on a target, such as an atom, as a result of DM scattering. In the absence of an observable signal, an exclusion region in the mass-cross-section parameter space is derived. To date, the strongest constraint comes from the LUX xenon-based experiment~\cite{Akerib:2016vxi}, constraining dark matter in the several GeV to 10 TeV mass range, with the DM-nucleon cross-section bounded to be below $\sim 10^{-46}\units{cm^2}$ at $\sim50$ GeV. An ongoing experimental  program will allow to push down this cross-section limit by two more orders of magnitude in the next decade~\cite{Aprile:2015uzo,Aalbers:2016jon}. However, none of the current ongoing experiments can probe significantly below the GeV mass scale.

\begin{figure*}
\centering
\includegraphics[width=0.9\textwidth]{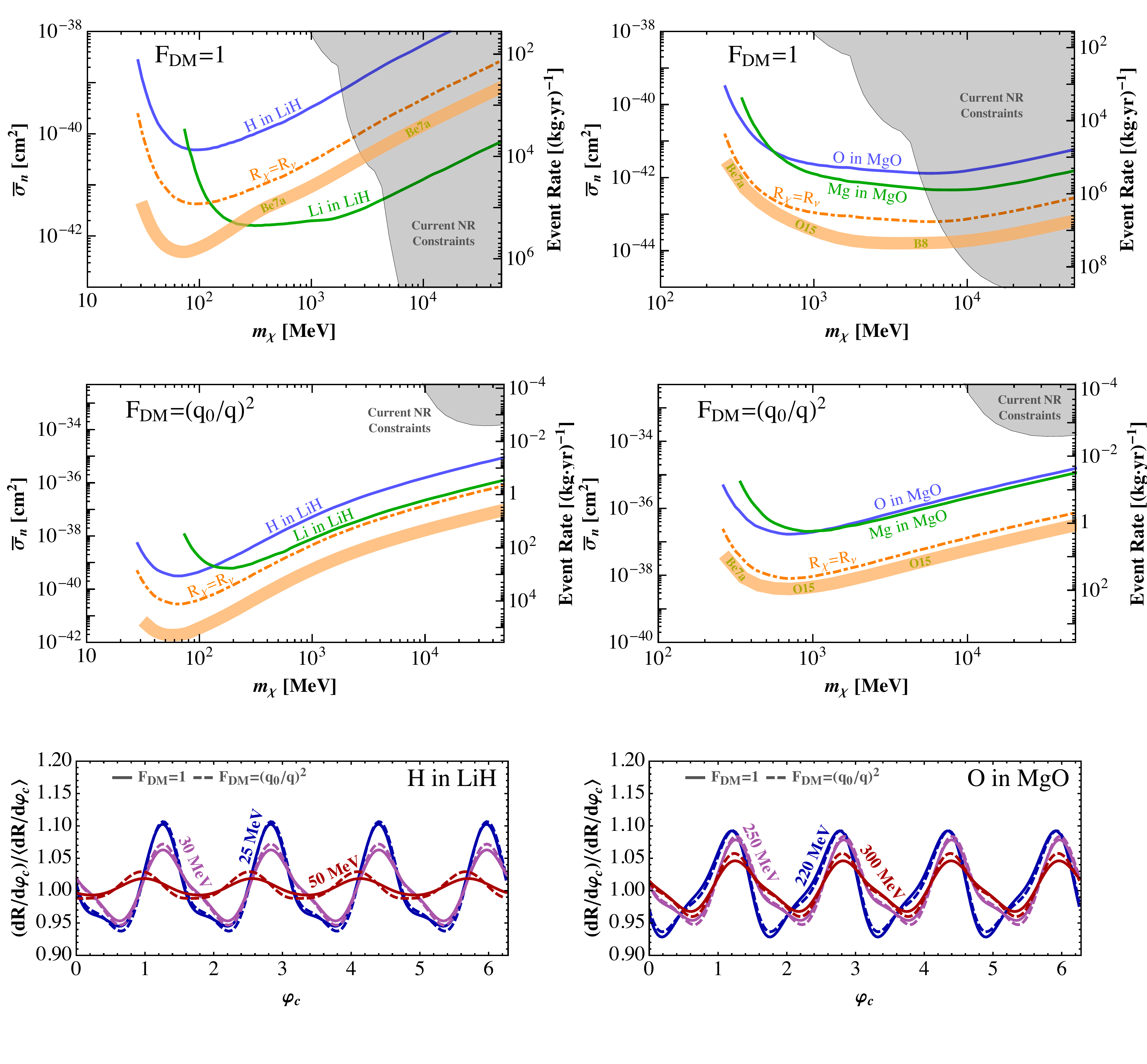} 
\caption{Potential cross section sensitivity for 1 kg$\cdot$year exposure ({\bf top} and {\bf center}) and daily modulation ({\bf bottom}) for the two example crystals considered in this study, assuming a background-free experiment and single defect sensitivity. Both crystals have CCs which have been studied in the literature (see text for details), LiH ({\bf left}) and MgO ({\bf right}). {\bf Top} and {\bf center}: The cross section sensitivity, given on the left axes, has been calculated for interactions with both types of nuclei within the crystals. The right axes correspond to the expected event rate per kg$\cdot$year assuming a DM-nucleon reference cross section of $\sigmap=10^{-37}\text{cm}^2$. {\bf Top} panels correspond to $\FDMq = 1$, {\bf center} panels correspond to $\FDMq = \left(q_0/q\right)^2$ with $q_0=100$ keV. The orange dotted-dashed curves correspond to the cross section for which one expects the rate of DM events to be equal to that of solar neutrino events for the lighter nucleus in each crystal (H in LiH and O in MgO). Below this line a dedicated neutrino background reduction will be necessary in order to detect DM. The thick orange curve corresponds to the prospective reach of a 100 kg$\cdot$year experiment for the same targets, following a dedicated neutrino reduction analysis. Black lines / gray shaded regions show nuclear recoil bounds from CRESST II, CDMSLite, SuperCDMS, and LUX. {\bf Bottom}: Daily modulation has been calculated for the lighter nucleus in each crystal (H in LiH and O in MgO). The modulation is presented as the differential rate normalized by its average for three DM masses, as a function of the angle between the crystal and the Earth's velocity in the galactic rest frame. Solid curves correspond to $\FDMq=1$ while dashed curves correspond to $\FDMq=\left(q_0/q\right)^2$. The latter show slightly larger modulation due to a stronger dependence on the minimal momentum transfer, except for DM masses very near threshold. Furthermore, lower masses correspond to larger modulations while the total rate is exponentially suppressed with decreasing mass. Modulation of order $\mathcal{O}(10)$'s\% is expected for these targets at these masses.}
\label{Fig:Rate_Mod_Plot}
\end{figure*}

The main shortcoming of current experiments arises  from the search for the elastic recoils of DM off nuclei. With an ${\cal O}($keV$)$ sensitivity to nuclear recoil energies, the suppression due to the  target mass does not allow to search for DM much below the GeV scale. Recently, it has been pointed out that searching for inelastic processes may allow to significantly lower the experimental threshold~\cite{Essig:2011nj}.
The specific realization of searching for bond-breaking phenomena has been studied in detail in~\cite{Essig:2016crl}.    In crystals, similar interactions  can induce detectable defects which may be searched for.
Here we explore the prospects of 
detecting the formation of defects known as color centers (CCs) following the dislocation of a nucleus within the crystal. 
These ${\cal O}(10)$ eV-threshold processes give rise to detectable signals, that should allow an experiment to explore an uncharted region in the parameter space of LDM as well as the low energy region of the solar neutrino spectrum. 

An exciting feature of near-threshold excitations in targets with intrinsic anisotropy, is the  directional dependence of the interaction rate.   Consequently, directional information in the form of a (sub-)daily modulation of the signal is expected, and would present as a striking signature of DM and a strong handle on background reduction.   

 The minimal cross sections which could be probed for a range of DM masses and for two example crystals considered in this study, with 1 kg$\cdot$year exposures, are presented in the top and center panels of Fig.~\ref{Fig:Rate_Mod_Plot}. Presented in the bottom panels of the figure is the expected daily modulation, i.e. the expected rates (normalized to their average) as a function of the orientation of the crystal with respect to the Earth's velocity (which, in turn, depends on the time of day and date). The momentum-transfer  dependent DM Form Factor (FF), $\FDM$, parameterizes  the microscopic physics of the DM-nucleon interaction. Other details of these results are given below. Here we stress the prospective two orders of magnitude improvement in DM mass sensitivity and the unique handle of daily modulation.

\section{Concept} 
\label{sec:Concept}

\begin{figure*}[t!]
\centering
\includegraphics[width=0.5\textwidth]{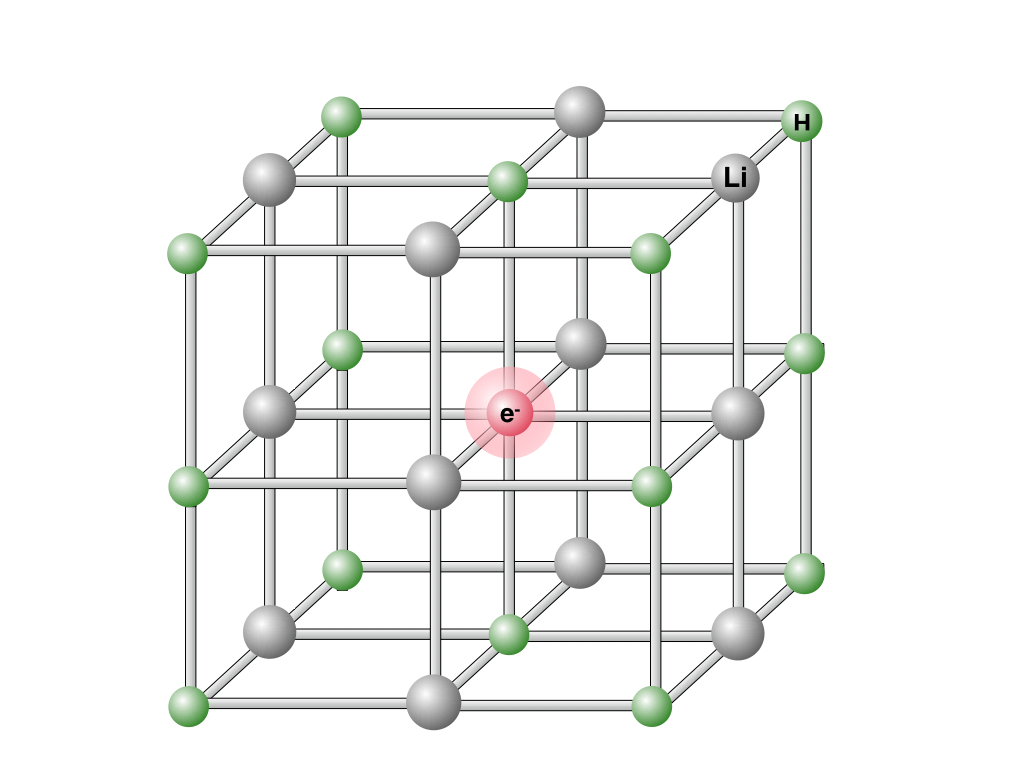} 
\caption{An illustration of an F-center in LiH. A singe H atom is removed from its global minimum and the charge deficit is occupied by a single electron whose energy levels translate to measurable fluorescent properties. 
\label{fig:CC}
}
\end{figure*}

Several requirements must be fulfilled in an experiment of the type discussed above, that aims at the discovery of LDM or other feebly-interacting particles:
\begin{itemize}
\item {\bf Low threshold}, typically below few tens of eV, in order to create a detectable defect.  
\item {\bf Background discrimination} which allows to differentiate between low-energy (signal) and high-energy (background) events, and between nuclear (signal) and electron (background) recoils, as well as between existing defects and newly formed ones.  
\item {\bf Detection capability} of defects. One possibility is the implementation of an enhancement mechanism such that any signal event can be detected multiple times.
\end{itemize}
Below, we discuss CCs and a possible experimental setup consistent with these principles.

CCs are atomic and electronic defects of various types in crystals, which produce optical absorption and emission bands in a spectral region to which the crystal is transparent~\cite{schulmancolor,Fowler}. These  defects occur in crystals such as alkali halides, ionic hydrides, alkaline earth fluorides and oxides. CCs are produced by $\gamma$ radiation, X-radiation or by energetic particles such as neutrons. One of the most well understood CCs, called F-centers, results from the vacancy of a negatively charged ion at a particular point within an ionic solid. This vacancy attracts and traps one or few electrons and this combination constitutes an F-center. A schematic view of an F-center in LiH is shown in Fig.~\ref{fig:CC}. 

In the context of this study, we propose the use of CC creation as a means to identify the scattering of a feebly-interacting particle (such as LDM or a neutrino) off an ion within the crystal. 
The result of the interaction is the transfer of kinetic energy and momentum to the ion. If the energy transfer is large enough, the ion will exit its global potential minimum and create a vacancy which may become a detectable CC.

Many of their properties make CCs ideal for the detection of a single dislocation event in a large volume. CCs live indefinitely at room temperatures and can be detected through their luminescence emission. Due to large oscillator strength, CC detection in a large volume is achievable if they are incorporated within a waveguide crystal rod. Many CCs exhibit low formation energies, of order few tens of eV, allowing for a low threshold experiment. Their nature may allow to differentiate between electronically- and nuclear-induced creation. 
Furthermore, 
CCs can be annealed to improve background rejection and effective experimental sensitivity. Finally, the crystal's intrinsic anisotropy allows for the study of signal intensity modulation, which can be used to positively identify signal events and to discriminate against various backgrounds, including the solar and atmospheric neutrino flux.

The study and use of 
defects in general and CCs in particular is an active field of research.  In recent years it  has  been exploited in many disciplines for the construction of sensitive detectors, (e.g. bolometers with Nitrogen-Vacancy centers in diamonds~\cite{2013PhR...528....1D})  as well as for the study of quantum computing (see e.g.~\cite{NatureQBits,PhysRevLett.107.220501}).  Existing techniques, however, use small-volume detection relying on the power of microscopy to probe single defects.  Here, we take a step forward and suggest large-volume detection of CCs, which requires alternative methods in order to study creation of single defects.



\section{Proposed Setup} 
\label{sec:Setup}


\begin{figure*}
\centering
\includegraphics[width=0.9\textwidth]{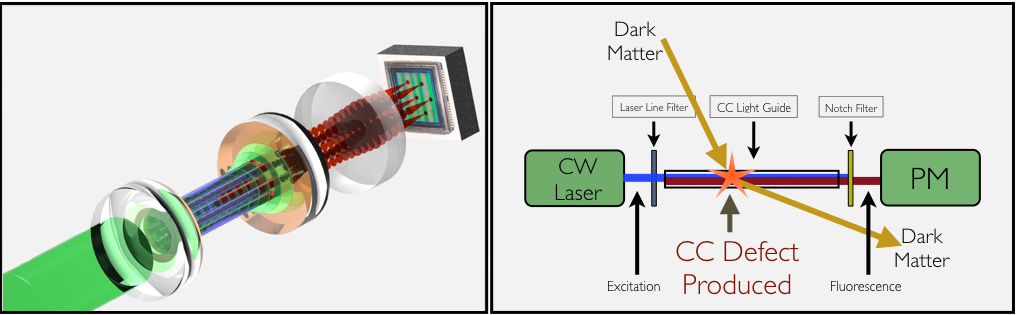} 
\caption{Schematic illustration of the envisioned setup for an experiment based on color-center excitations in crystals. A light guide composed of a crystal such as Lithium Hydride or Magnesium Oxide acts as the target material. A CW laser with an excitation frequency tuned to the color center is then used to induce fluorescence emissions, which can be detected on the other side with  photomultipliers or SCMOS cameras. }
\label{fig:setup}
\end{figure*}

An interacting particle passing through a crystal of the type discussed above may form one or few CCs that can be identified via their absorption and emission properties. We envision a setup that will follow a macroscopic amount of material in pure crystal form, performing a counting experiment of defects of one or more kinds with adequate spatial and temporal resolutions. The goal of the experiment is to identify  the production of a single or very few CC sites, over the background of existing sites and instrumental noise.  An illustration of such a setup is shown in Fig.~\ref{fig:setup}. The crystal detector is composed of an array of light-guiding rods, imaged by a high efficiency, high resolution array of photon counters such as SiPM, or SCMOS cameras. The exciting beam is enhanced inside the rods, while a notch filter blocks the excitation and allows the luminescence to reach the photon counter array.  The use of separate rods, which are imaged individually, provides smaller imaged volumes. This translates to lower background for a single CC generation event and also allows for natural spatial resolution and multiple scattering identification.

Quantitatively,  an excitation beam of a $\sim1$~W CW laser  inside the cavity, built such that the intensity inside the rods is tenfold higher, leads to an estimated $10^5$ photons detected per single CC site for a 10-15 min exposure\footnote{Assuming 100\% quantum efficiency for light emission, the number of fluorescence photons detected, N, over a photon counting period, $\Delta t$, is given by,
$N = R \times \Delta t \times \eta_{\rm det}  \times \Omega_{\rm collect}$.  Here $R$ is the CC excitation rate, $\eta_{\rm det}$ is the quantum efficiency of the light detector, and $\Omega_{\rm collect}$ is the light collection efficiency. For typical numbers of the system (photon flux, $\Phi=10^{21}$ $\text{cm}^{-2}\text{ s}^{-1}$, absorption cross-section, $\sigma=10^{-17}\text{cm}^2$, $\Omega_{\rm collect}=0.1$, $\eta_{\rm det}=0.5$ and $\Delta t=10^3 \text{ s}$) the count of photons equals $5\times10^5$.}. This relies on   realistic estimates of intensities, lifetimes, detection efficiency and cross sections for excitation.  A detection limit of about one site in $10^5$ background sites in a single rod during that time period is thus inferred (at 1$\sigma$). The choice of rod size and aspect ratio, as well as exposure details, will be based on the achievable defect cleanliness and exact properties of the spectroscopy. Consequently, it is expected that the number of  CCs produced in a measurement window can be estimated with high accuracy. The outcome of the experiment will be the  differential rate of multiple CC production, where the number of produced CCs  is plausibly correlated with the deposited energy in a given scattering event. Experimental and theoretical studies of such correlations are in need and will be presented in future publications.

Multiple pairs of excitation and emission wavelengths can be used, with a more complex optical setup, which would
provide another handle on the microscopic interaction.  For example, sensitivity to different CCs produced by the displacement of different nuclei would allow to better identify the DM mass and DM-nucleon form-factor (related to the mediation scheme of DM to the visible sector), thereby enhancing the sensitivity to the DM parameter space.  A better handle on recoil energy spectrum  may also be gained via measurement of production of different CCs with varying thresholds, within a given crystal.


Intrinsic properties of CC production and crystal symmetries lead to a dependence of the threshold energy on the direction of recoil relative to the crystal axis. This can be exploited for directional information and thus for background reduction. Measurement of any modulation of the signal may also contribute to the knowledge regarding the properties of the LDM. Specifically, the modulation properties depend on the absolute interaction rate, the mass and the interaction type of the LDM particle (detailed discussion below). 

\subsection{Backgrounds}
A major concern with  rare event detectors is the background. We divide these into three major categories:

{\it Existing defects}: Crystals, even in the purest form, contain defects of many kinds. Some are chemical impurities while others are structural defects from production or past radiation interactions. These can potentially mimic the desired signal. The choice of specific wavelengths for emission and absorption, filters most of this background noise. However, as commercial manufacturing typically does not provide guaranties for defect concentration lower than 10$^{-9}$, it is expected that the concentration of relevant sites  in virgin crystals will be much higher than that required for a running DM detector. A  solution is annealing of the crystals in a low background environment by gradually heating to high (up to $\sim$1000$^\circ$C) temperature. This  allows for the curing of a large fraction of these defects~\cite{Davidson}.

{\it Radiogenics}: Current DM direct detection experiments constantly battle   radioactive background sources which in many cases are the limiting factor for the physics reach. For CC production at low energies, this focus is somewhat different since direct production of a single, relevant CC  by incident $\alpha, \beta$ or $\gamma$ radiation is strongly suppressed in the bulk, as the resulting high energy deposit can easily be detected through the almost simultaneous production of a large number of defects of various kinds. An exception is the nuclear, coherent Thomson scattering of $\gamma$ on nuclei. However, this is expected to be extremely suppressed in comparison to Compton scattering \cite{PhysRevD.95.021301} and is unlikely to present an experimental challenge for the exposures discussed here. This leaves neutrons as the single most challenging radiogenic background. Neutrons can be dealt with by hydrogen rich passive shielding and possibly by an active neutron veto, e.g. a boronated scintillator surrounding the detector enclosure. Similar geometry experimental setups such as CDMS-II \cite{Ahmed:2008eu} (121 kg$\cdot$day) have succeeded in efficiently removing this type of background. 
 
{\it Cosmogenics}: Solar neutrinos scattering off nuclei are an irreducible background. However, these exhibit (a) {\it very low interaction rates} which will only become relevant at exposures above $\sim10-100$~kg$\cdot$years, integrated over the entire energy spectrum (details below) and (b) {\it a first ever detection of neutrino-nucleus coherent scattering for most of the solar spectrum}. Apart from neutrinos, muons arriving at an underground lab can pass through the detector or induce showers of energetic particles. These will have to be screened by an external active muon veto as is customary in many other  DM direct detection experiments, e.g. \cite{Ahmed:2008eu}.

\section{Physics Prospects}
\label{sec:Physics}



DM-nucleus or neutrino-nucleus scattering can be thought of classically, as a two-step process. Initially, the weakly interacting particle scatters with the nucleus, transferring some momentum $\mathbf{q}$, followed by the subsequent escape of the nucleus from its potential well.  
The first step occurs on a timescale much shorter than the timescale for the escape of the nucleus. Therefore, the interaction itself can be decoupled from the rest of the process and its outcome can be thought of as a nucleus sitting at the origin with some initial momentum $\mathbf{q}$. 
Since this is a many body system, as the nucleus moves, it scans the potential in a directionally dependent manner, losing energy to dissipation as it interacts with the time dependent potential. This process depends both on the direction and the absolute value of $\mathbf{q}$ and results 
in either the dissociation of the nucleus from the initial minimum of the potential or relaxation back to the initial state. For the crystals we have in mind, the entire process is well approximated by a classical solution~\cite{Essig:2016crl}\footnote{For the lightest target we consider, quantum effects are expected to change the DM mass threshold of the experiment by $\lesssim15$\%. For the heavier targets the correction is further suppressed.}, and therefore in what follows we calculate the scattering rates treating the system classically.

\subsection{Dark Matter Scattering}

The DM-nucleus scattering rate is given by (see Appendix for details),
\begin{eqnarray}
\frac{\partial R}{\partial \varphi_c} & = & f_{PN}^2 N_T \frac{\rho_\chi}{m_\chi} \frac{\sigmap}{16 \pi^2 \mu_{\chi n}^2} \int dq^2 \int d^3v \int d\varphi^{\prime\prime}_{q} \label{eq:Clas_Rate} \\
& & \times \FDMsq \Theta \left(q - q_{\rm min}(q,\varphi^{\prime\prime}_{q},\mathbf{v},\varphi_c) \right) \frac{f(\mathbf{v})}{v}, \nonumber \
\end{eqnarray}
where  $f_{PN} = \left[ f_PZ+f_N(A-Z) \right]$ is a coherence factor which depends on the atomic ($Z$) and mass ($A$) numbers of the target and the coupling strengths of the interacting DM to the proton ($f_P$) and neutron ($f_N$). 
Furthermore, $N_T$ is the number of target nuclei, $\rho_\chi$ and $m_\chi$ are the DM density and mass respectively, $\mu_{\chi n}$ is the reduced mass of the DM-nucleon system and $f(\mathbf{v})$ is the velocity profile of DM in the Milky-Way halo (see e.g. \cite{Lewin:1995rx}) where we adopt $v_0 = 230 \text{ km/s}$, $v_{\rm Earth} = 240 \text{ km/s}$, and $v_{\rm esc} = 600 \text{ km/s}$. The function $\FDM$ is the DM form factor which encodes information about the DM-target interaction (see discussion in~\cite{Essig:2016crl}) such that $\sigmap\cdot\FDMsq$ is the DM-nucleon interaction cross section. With this definition, $\sigmap$ is the DM-nucleon interaction cross section for $f_P=f_N=1$, at some reference momentum transfer value, $q_0$, which we take to be $q_0=100$~keV.
 
The function $q_{\rm min}(q,\varphi^{\prime\prime}_{q},\mathbf{v},\varphi_c)$ is the minimal value of momentum transfer which is required in order to dissociate the nucleus in a given direction. Its target-dependent values depend on the direction of $\mathbf{q}$ with respect to the orientation of the crystal. This direction in turn depends on the kinematics of the interaction and the orientation of the crystal with respect to the incoming DM wind. The angle $\varphi_c$ defines this orientation with respect to the Earth's velocity in the galactic rest frame. The angle between the velocity vector of the incoming DM, $\mathbf{v}$, and the momentum transfer, $\mathbf{q}$, is set by their absolute values, $v$ and $q$ respectively. The momentum transfer is still free to rotate around the axis defined by $\mathbf{v}$, where this angle of rotation is denoted by  $\varphi_q^{\prime\prime}$. Thus, the direction of momentum transfer is a function of the variables $q$, $\varphi^{\prime\prime}_{q}$, $\mathbf{v}$ and $\varphi_c$. After integration over the first three of these, the final result for the interaction rate depends on $\varphi_c$. This presents itself in the form of a daily modulation of the rate. A detailed discussion can be found in the Appendix.

We note that the rate given in Eq.~(\ref{eq:Clas_Rate}) is that for creating a dislocated nucleus within the crystal and not for the creation of a specific defect type. For any target, there are expected to be multiple possible defects created following a momentum kick to a nucleus. The creation of a specific defect depends on the momentum transfer vector (and possibly on other effects), thus, since the rate above has been calculated after integration over the entire momentum spectrum, it should be thought of as the integrated rate for the creation of all possible allowed defects.

To demonstrate our results we consider two crystals, MgO and LiH, with CCs which have been studied in the litariture~\cite{Kotomin19981,0022-3719-18-27-015}. For both crystals, we consider the dislocation of both ions within the crystals. These examples give intuition for expectations from a wide range of masses and potentials. Both crystals exhibit average binding energies of order 10's of eV and target masses which range from few to $\mathcal{O}(10)$ GeV. The optimal choice of crystal depends on the mass and cross section of the LDM. Light target nuclei reduce the DM mass threshold while heavier targets amplify the interaction cross section 
due to a coherent enhancement of the interaction rate.
Small binding energies reduce the DM mass threshold. For these reasons a crystal such as LiH is expected to be ideal for detection of extremely light DM down to $\sim 20$ MeV. If DM is heavier, other crystals may be more suitable in order to enhance the cross section.  Studying F-centers induced by the dislocation of both atoms within a crystal thus greatly broadens the sensitivity to the DM parameter space. 

 For each crystal, we model the function $q_{\rm min}$ as,
\beq
q_{\rm min}(\Omega^\prime_q) = \sqrt{2 m_1 \DE \left[ 1 + a \sin{(n_\theta \theta^\prime_q)} \cos{(n_\varphi \varphi^\prime_q)} \right]},
\eeq
where $n_\theta, n_\varphi$ parametrize the periodicity of the crystal and $a$ is the amplitude of the variation of the binding energy. The value $m_1$ is the mass of the nucleus undergoing the interaction and $\DE$ is the mean binding energy of the $m_1$ ion to its site. The solid angle $\Omega^\prime_q$ corresponds to the direction of $\mathbf{q}$ and is evaluated in a coordinate system which is fixed to the orientation of the crystal target. As discussed above, this angle is a function of $q$, $\varphi^{\prime\prime}_{q}$, $\mathbf{v}$ and $\varphi_c$ (see detailed discussion in the Appendix). For both crystals considered in this study we take $n_\theta = n_\varphi =4$, which is the case for a cubic crystal.

Most defects have been studied for only a small number of directions of momentum transfer to the nucleus. We take conservative estimates of $\DE$ and of $a$ such that the minimal values of $q_{\rm min}^2/2m_1$ are always larger than the minimal values of binding energy found in the literature. Thus, our results are at best underestimations of the rate. Specifically, for MgO we choose $\DE$ as the mean value of binding energies found in the literature~\cite{Kotomin19981} and choose the value of $a$ such that the minimal $q_{\rm min}$ is always larger than the minimal value which has been measured. For LiH, theoretical values of formation energies have been calculated for the creation of F-center defects as well as Schottky defects, where the values for the latter are in agreement with experiment~\cite{0022-3719-18-27-015}. Here we assume the same approximate value of $a$ as in MgO and choose $\DE$ such that the minimal value of $q_{\rm min}$ is twice as large as the calculated formation energy for F-centers (this accounts for a possibly large potential barrier). All values are presented in Table~\ref{Tab:DeltaEb}. 

\begin{table}[t!]
 \begin{center}
\begin{tabular}{| c | c | c | c |}
\hline
\multirow{2}{*}{MgO} 
 & $\Delta E_{B \text{,Mg}}$ =  55 eV & $a_{\rm Mg} = 0.4$ & \cite{Kotomin19981}  \\
 & $\Delta E_{B \text{,O}}$ =  50 eV & $a_{\rm O} = 0.4$ & \cite{Kotomin19981}  \\
 \hline
 \multirow{2}{*}{LiH} 
 & $\Delta E_{B \text{,Li}}$ =  9 eV & $a_{\rm Li} = 0.4$ & \cite{0022-3719-18-27-015}  \\
 & $\Delta E_{B \text{,H}}$ =  9 eV & $a_{\rm H} = 0.4$ & \cite{0022-3719-18-27-015}  \\
 \hline
\end{tabular}
 \end{center}
\caption{Values taken for the mean binding energies and modulation parameters for the crystals considered in this study.\label{Tab:DeltaEb}}
\end{table}

In the top and center panels of Fig.~\ref{Fig:Rate_Mod_Plot}  we present the expected average cross section reach assuming zero background, and the event rate assuming a reference cross section $\sigmap = 10^{-37} \text{ cm}^2$, for the two reference crystals, with a 1 kg$\cdot$year exposure and $\FDMq=1$ or $\FDMq=\left(q_0/q\right)^2$
 (see \cite{Essig:2016crl} for motivation for these choices). We calculate the rate of site formation and assume that electron occupation is spontaneous and immediate. We also assume the (analysis related) maximal value of momentum transfer to be~$q_{\rm max} = 10 \sqrt{2 m_1 \DE (1+a)}$. We find that with 1 kg$\cdot$year, an LiH crystal can potentially probe DM-nucleon cross sections down to $\sim10^{-42}\text{ cm}^2$ and DM masses down to $\sim20$ MeV. An MgO crystal can potentially probe DM-nucleon cross sections down to $\sim5\cdot10^{-43}\text{ cm}^2$ and DM masses down to $\sim200$ MeV. The curves for $\FDMq=\left(q_0/q\right)^2$ are somewhat less sensitive than for $\FDMq = 1$, however this is partly due to the fact that we have chosen $q_0 = 100$ keV for all targets, while this is an underestimation of the typical momentum transfer for the heavier targets which probe larger DM masses. Finally, the dotted-dashed orange curves in these panels correspond to the cross section for which solar neutrino reduction becomes important, defined as the cross section for which the DM rate equals the total rate of solar neutrino events. The thick orange curves correspond to the projected sensitivity after solar neutrino background reduction for a $100\text{ kg}\cdot\text{year}$ exposure (see discussion below). Also shown in the figures in shaded grey are the bounds derived from current nuclear recoil experiments, namely CRESST II~\cite{Angloher:2014myn}, CDMSLite~\cite{Agnese:2013jaa}, SuperCDMS~\cite{Agnese:2014aze}, and LUX~\cite{Akerib:2015rjg}.
 
 \begin{figure*}[t!]
\centering
\vspace{5pt}
\includegraphics[width=0.98\textwidth]{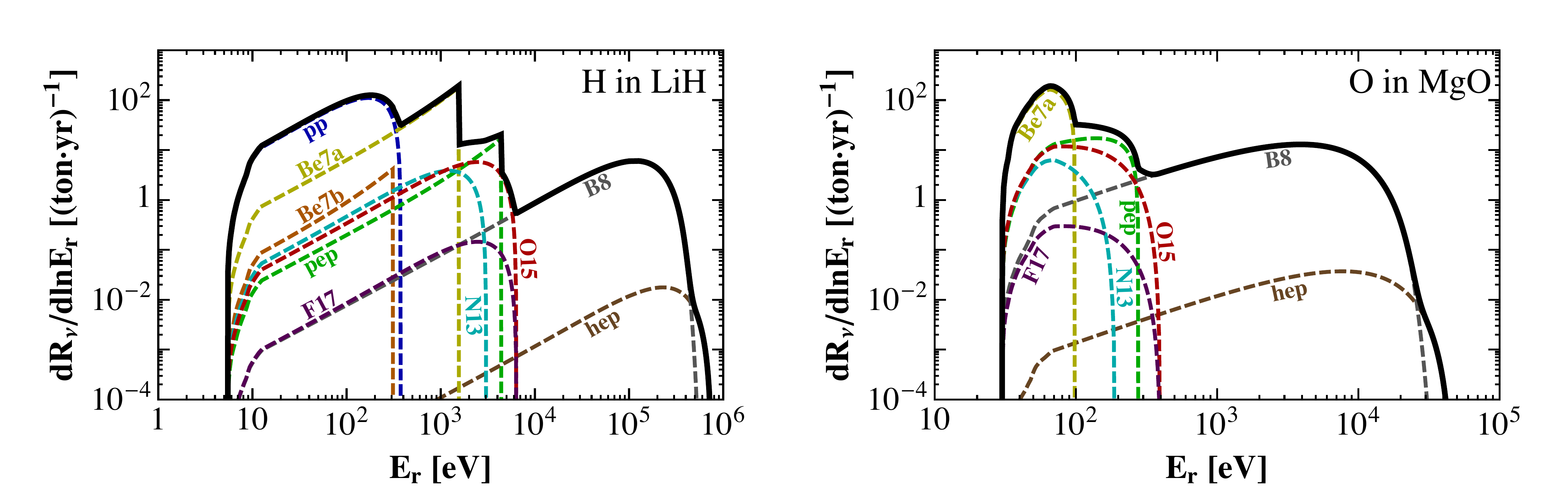} 
\caption{The solar neutrino recoil energy spectrum for H in LiH ({\bf left}) and for O in MgO ({\bf right}) for a 1 ton$\cdot$year experiment. Each labelled colored curve marks the contribution from one of nine families of solar neutrinos. Light targets pronounce the low energy part of the spectrum, namely the pp neutrinos. The total number of events is $\sim380 \text{ events}/\text{ton}/\text{year}$ for H in LiH (of these $\sim190$ events result from the pp spectrum) and $\sim180 \text{ events}/\text{ton}/\text{year}$ events for O in MgO. The result for H has been calculated with both the spin independent and the spin dependent contribution to the neutrino-nucleus cross section.
\label{Fig:dRnudlnEr}
}
\end{figure*}
 
In the bottom panels of Fig.~\ref{Fig:Rate_Mod_Plot}, we present daily modulation of the rate for interactions with H in LiH and with O in MgO. Plotted in the figure is the differential rate $\partial R/\partial \varphi_c$, normalized to the mean differential rate, as a function of $\varphi_c$, which is proportional to the time of day. Solid curves correspond to the trivial DM FF, $\FDMq = 1$ while dashed curves correspond to $\FDMq = \left(q_0/q\right)^2$. For a given DM mass, modulation occurs mainly in the low energy part of the recoil energy spectrum. Thus, DM masses whose spectrum is dominated by low energies (of order the binding energy) present larger modulation. Also, for $\FDMq = \left(q_0/q\right)^2$, modulation is slightly larger than for $\FDMq = 1$, since for the latter case the result is dominated by the maximal momentum transfer in the integral while for the non-trivial DM FF, the integral is dominated by the minimal allowed, $\varphi_c$ dependent, momentum transfer. This effect becomes less important close to mass threshold (details given in the Appendix). It should also be noted that as the DM mass is lowered, the total rate becomes exponentially suppressed. Therefore, larger modulations correspond also to lower total rates. The periodicity of the modulation depends on the periodicity of the binding potential as can be seen in the figure.



\subsection{Neutrino Scattering}

Neglecting quantum corrections (i.e. the target form factor) the inelastic spin-independent\footnote{The spin-independent neutrino-nucleus cross section is a good approximation for atoms with large atomic numbers. Results shown for H have been calculated using both the spin-dependent and spin-independent contributions. Details can be found in Appendix B of \cite{Essig:2016crl}.} neutrino-nucleus cross section is,
\beq
\frac{\partial\sigma}{\partial E_r \partial \varphi_q^{\prime\prime}} = G^2 \frac{M}{8\pi^2} [Z(4 \sin^2\theta_W-1) + N]^2 \left(1-\frac{m_1 E_r}{2E_\nu^2}\right),
\eeq
and the differential dissociation rate is,
\begin{eqnarray}
\frac{\partial R_\nu}{\partial \ln{E_r}\partial\varphi_c\partial\varphi_s} & = & \frac{N_T m_1 E_r}{4\pi^2} \int d E_\nu \int d \varphi^{\prime\prime}_q \frac{\partial F}{\partial E_\nu} \frac{\partial\sigma(E_r,E_\nu)}{\partial E_r \partial \varphi_q^{\prime\prime}} \nonumber \\
& & \times \Theta \left( E_r-E_{r,{\rm min}}[\Omega_q^\prime (q,E_\nu,\varphi_q^{\prime\prime},\varphi_s,\varphi_c)] \right), \nonumber \\
&& \
\label{eq:dRnudlnEr}
\end{eqnarray}
where $E_\nu$ is the incoming neutrino energy, $\partial F/\partial E_\nu$ is the differential neutrino flux, and $N_T$ is the number of targets in the system as before. The angle $\varphi_s$ corresponds to the direction of the incoming neutrino flux with respect to the target (see Appendix for details) and $\varphi_c$ is as above. For the case of Solar neutrinos, the angle $\varphi_s$ corresponds to the direction of the Sun, and the differential rate will depend on both variables $\varphi_s, \varphi_c$. Daily modulation for the case of Solar neutrinos is small, since these typically have energies (and corresponding energy transfers) much larger than the binding energies of nuclei within the lattice. The small daily modulation that does occur will have an annual phase change which depends on $\varphi_s$.

In Fig.~\ref{Fig:dRnudlnEr} we present the predicted solar neutrino spectra for H in LiH and for O in MgO. The solar pp spectrum is pronounced for the lighter target and is the main contribution to the low energy spectrum. Thus, a target material such as LiH may allow for a first measurement of the pp spectrum via coherent nuclear scattering. For an exposure of 1 ton$\cdot$year, the total number of events is $\sim380 \text{ events}/\text{ton}/\text{year}$ for H in LiH (with $\sim190$ of these events resulting from the pp-neutrino spectrum) and $\sim180 \text{ events}/\text{ton}/\text{year}$ events for O in MgO. 

For the DM detection discussed above, solar neutrino reduction becomes important approximately at DM-nucleon cross sections for which the number of DM events is equal to the number of neutrino events. The orange dotted-dashed curves in the top and center panels of Fig.~\ref{Fig:Rate_Mod_Plot} corresponds to this cross section (for the lighter nucleus in each crystal). This curve is exposure independent since both the DM and neutrino rates scale with exposure. Approximately at this line and below, a dedicated background reduction is necessary in order to claim a potential DM discovery. Following~\cite{Ruppin:2014bra}, we have calculated the maximum reach for the DM-nucleon cross section for an experiment with a 100 kg$\cdot$year exposure, shown by the thick orange curves in the top and center panels of Fig.~\ref{Fig:Rate_Mod_Plot}. Features in these curves correspond to DM masses whose spectra happen to mimic a certain contribution to the solar neutrino spectra. For these masses, the uncertainty in the event rate may become dominated by the uncertainty in the neutrino flux and the reach begins to saturate with growing exposure (the neutrino family which dominates the uncertainty at each DM mass is shown within the thick orange curve in the figure). The so-called neutrino floor for this class of experiments is reached only at extremely large exposures and is therefore not presented in the figures.

\section{Conclusions} 
\label{sec:Conclusions}
We have proposed a new ultra-low threshold technique for the direct detection of DM a light as $\sim20$ MeV. Our proposed setup will also be sensitive to the low energy spectrum of solar neutrinos, including the pp-neutrino spectrum. If measured, this would be the first ever measurement of coherent nuclear scattering of the pp spectrum. We expect the detection of single defects in a macroscopic bulk of target material to be possible, and that the relevant backgrounds to be controllable. Such an experiment also features directional information which present in the form of daily modulation of the signal. Measurement of the modulation will provide a handle on background reduction and possibly information on the DM parameters. For an exposure of 1 kg$\cdot$year, one could probe DM-nucleus cross-sections down to $\sim5\cdot10^{-43} \text{ cm}^2$ for a trivial DM form factor, and masses in the $\mathcal{O}(10)-\mathcal{O}(100)$ MeV range, depending on the target material. For this exposure, the number of solar neutrino events is expected to be $\lesssim1$, so that detection of solar neutrinos requires larger exposures.

{\bf Note Added.}  While this paper was being completed, ref.~\cite{Kadribasic:2017obi} appeared, where a similar study of the daily modulation in crystals was discussed. 

{\it\bf Acknowledgments:}
We  thank Rouven Essig and Jeremy Mardon for many useful discussions.   
We also thank Eitam Vinograd for creating the left of Fig.~\ref{fig:setup}.   This work is supported in part by the PAZI foundation.  RB, OS and TV are supported by the I-CORE Program of the Planning Budgeting Committee and the Israel Science Foundation (grant No. 1937/12).
TV and OS are also supported in part by the European Research Council (ERC) under the EU Horizon 2020 Programme (ERC-CoG-2015 - Proposal n. 682676 LDMThExp).  TV is further supported by the German-Israeli Foundation (grant No. I-1283- 303.7/2014).
OS is further supported by the Clore Foundation. RB is the incumbent of the Arye and Ido Dissentshik Career Development Chair.

\section{Appendix}
\label{sec:App_1}

\begin{figure*}[t!]
\centering
\vspace{5pt}
\includegraphics[width=0.25\textwidth]{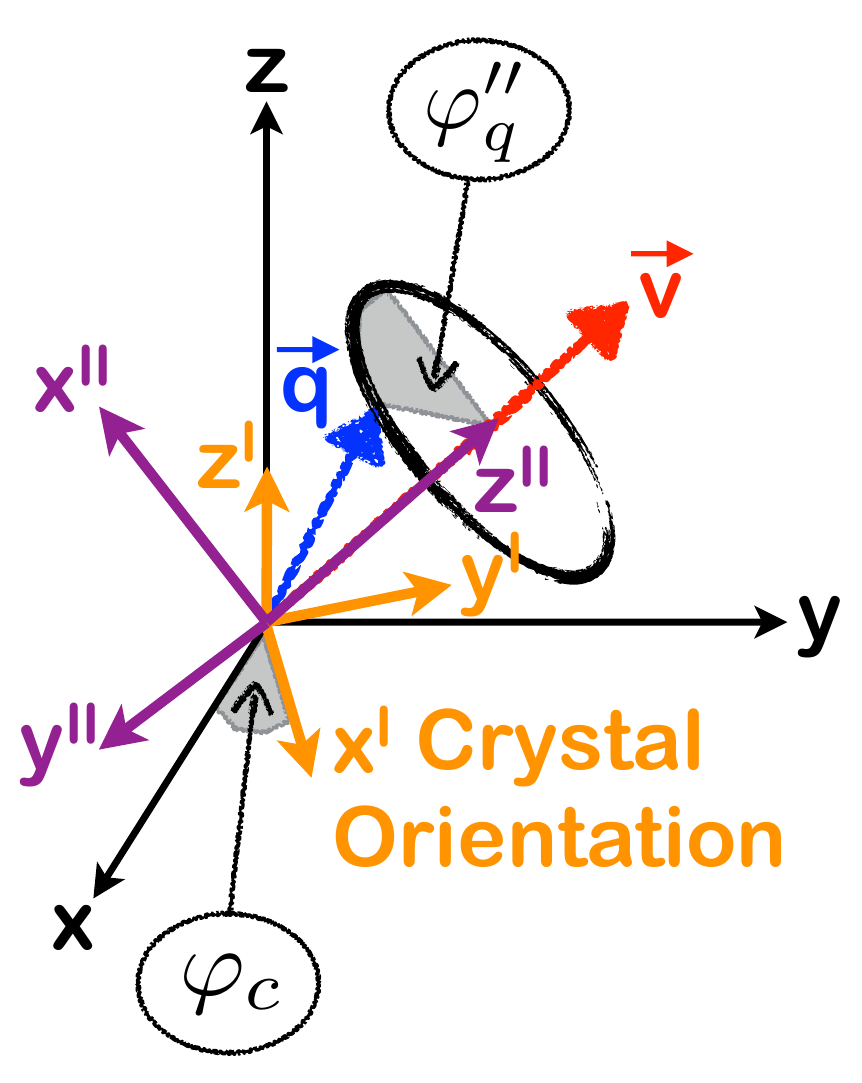} 
\caption{Coordinate systems defined for calculation of differential rates and modulation. \textbf{Black}: \textit{The Earth Frame} ($x,y,z$) whose $z$ axis is parallel to the Earth's rotation axis and is fixed to the galactic rest frame. \textbf{Orange}: \textit{The Lab Frame} ($x^\prime,y^\prime,z^\prime$) whose $z^\prime$ axis is parallel to $z$ and the $x^\prime$ axis is fixed to some orientation within the crystal target. \textbf{Purple}: \textit{The DM Velocity Frame} ($x^{\prime\prime},y^{\prime\prime},z^{\prime\prime}$) whose $z^{\prime\prime}$ axis is parallel to the direction of the initial DM velocity. The $x^\prime$ axis (the crystal orientation) rotates around the $z$ axis with an angle $\varphi_c$ which has a period of one day. The angle between $\mathbf{q}$ and $\mathbf{v}$ depends on their absolute values. With this angle fixed, $\mathbf{q}$ is still free to rotate around the $\mathbf{v}$ direction, defined by the angle $\varphi_q^{\prime\prime}$, which must be integrated over in order to calculate the interaction rate.
\label{Fig:coordinates}}
\end{figure*}

The physics involved in scattering of a weakly interacting particle with a nucleus within a crystal lattice is somewhat similar to that described in~\cite{Essig:2016crl}. However, for the case of a scattering within a lattice, since the potential is not spherically symmetric, the outgoing nucleus follows a non-trivial path as it exits its binding potential, losing part of its energy to dissipation.
The corresponding path depends  on the direction of the momentum transfer and on its absolute value. Therefore, calculating the rate as a function of the binding potential is inconvenient. An alternative is to discuss the minimal recoil energy or equivalently the minimal momentum transfer, which is a function of the direction of the momentum transfer only,
\beq
E_{\rm r}^{\rm min}(\Omega^\prime_q) = \frac{q_{\rm min}^2(\Omega^\prime_q)}{2 m_1},
\eeq
where $\Omega^\prime_q$ is the direction of $\mathbf{q}$ in the rest frame of the target crystal (this frame is denoted with a single prime). The directionally-dependent minimal energy follows from calculating the classical path for a dissociated nucleon initially at the potential minimum, after receiving a momentum kick. Each classical path takes into account the dissipative multi-scattering process and can be mapped to such a minimal recoil energy.

As was shown in~\cite{Essig:2016crl}, the dissociation rate depends only mildly on the form of the binding potential and is almost solely dependent on the binding energy, as long as the initial wavefunction is sufficiently localized around the origin. 
Furthermore, quantum corrections become small when,
\beq
\frac{\Delta p}{q_{\rm min}} \ll 1,
\eeq
where $\Delta p$ is the typical momentum spread in the nucleus' initial state. Since the typical uncertainty in the position of the nucleus is of order the Bohr radius, it follows that $\Delta p \sim \alpha m_{\rm e}$. Therefore, for most targets a classical approach is sufficient.

\subsection{DM-Nucleus Scattering Rate for Crystal Targets}
Classically, the average differential DM-nucleus interaction cross-section is,
\beq
\left<\frac{\partial \sigma v}{\partial q^2 \partial \varphi^{\prime\prime}_{q}} \right> = f_{PN}^2 \frac{\sigmap}{8 \pi \mu_{\chi n}^2 v} \FDMsq,
\label{eq:Class dsigma_dq2}
\eeq
where $f_{PN} = \left[ f_PZ+f_N(A-Z) \right]$ with A and Z the mass and atomic numbers of $m_1$, $f_P$ ($f_N$) is the coupling strength to the proton (neutron), $\mu_{\chi n}$ is the reduced-mass of the DM-nucleon system and $\FDM$ is the DM form factor~\cite{Essig:2016crl}. Setting the values of $q$ and $v$ sets the angle between them (this follows from energy conservation) according to,
\beq
c_\theta = \frac{q}{2 m_\chi v}.
\eeq
The angle $\varphi^{\prime\prime}_{q}$ defines the rotation of $\mathbf{q}$ around $\mathbf{v}$ with $c_\theta$ fixed.

In order to calculate the differential interaction rate, one must integrate over all possible values of the DM velocity vector, $\mathbf{v}$, and over the angle $\varphi^{\prime\prime}_{q}$. This should be done in a coordinate system which is fixed to the rotation axis of the Earth, since this direction is fixed throughout the day. It is convenient to define three coordinate systems, illustrated in Fig.~\ref{Fig:coordinates}, in order to perform the calculation:
\begin{itemize}
\item \textit{The Earth Frame} whose $z$ axis is parallel to the Earth's rotation axis and is fixed to the galactic rest frame. This system is denoted ($x,y,z$) and is shown in black in the figure.
\item \textit{The Lab Frame} whose $z^\prime$ axis is parallel to $z$ and the $x^\prime$ axis is fixed to some orientation within the crystal target. This system is denoted ($x^\prime,y^\prime,z^\prime$) and is shown in orange in the figure.
\item \textit{The DM Velocity Frame} whose $z^{\prime\prime}$ axis is parallel to the direction of the initial DM velocity. This system is denoted ($x^{\prime\prime},y^{\prime\prime},z^{\prime\prime}$) and is shown in purple in the figure.
\end{itemize}
With these definitions, the $x^\prime$ axis (the crystal orientation) rotates around the $z$ axis with an angle $\varphi_c$ which has a period of one day, the angle $\varphi_q^{\prime\prime}$ is well defined in the DM Velocity Frame (see Fig.~\ref{Fig:coordinates}) and the DM velocity profile has a simple form in the Earth Frame.

The kinematically allowed range of momentum transfer is,
\beq
q_{\rm min}(\Omega^\prime_q) \leq q \leq 2 \mu_{\chi 1} v.
\eeq
The second inequality sets either an upper limit on $q$ or a lower limit on $v$, while the first inequality is a lower limit on $q$ defined by the shape of the binding potential and which depends on the direction of $\mathbf{q}$ in the Lab Frame, $\Omega^\prime_q$. This direction is a function of the variables $\Omega^\prime_q=\Omega^\prime_q(q,\mathbf{v},\varphi^{\prime\prime}_{q},\varphi_c)$. Therefore, after integrating over $\mathbf{v}$ and $\varphi_q^{\prime\prime}$, the differential scattering rate has an explicit $\varphi_c$ dependence and is given by,
\begin{eqnarray}
\frac{\partial R}{\partial q^2\partial\varphi_c} & = & f_{PN}^2 N_T \frac{\rho_\chi}{m_\chi} \frac{\sigmap}{16 \pi^2 \mu_{\chi n}^2} \int d^3v \int d\varphi^{\prime\prime}_{q} \label{eq:Clas_Rate_app} \\
& & \times \FDMsq \Theta \left(q - q_{\rm min}(q,\mathbf{v},\varphi^{\prime\prime}_{q},\varphi_c)) \right) \frac{f(\mathbf{v})}{v}, \nonumber \
\end{eqnarray}
where $f(\mathbf{v})$ is the velocity distribution of DM~\cite{Lewin:1995rx} and the integral over $v$ has a minimum at $v_{\rm min}(q) = q/2 \mu_{\chi 1}$.

The amplitude of the daily modulation of the signal can be understood by noticing that the total rate has a dependence which scales differently for the two DM FFs we have considered. Namely, the scaling is,
\beq
R \propto \frac{A^2}{m_\chi \mu_{\chi 1}^2 m_1} \left\{
\begin{array}{lll}
q_{\rm max}^2 - q_{\rm min}^2 & , & \FDMq = 1 \\
\frac{q_0^4}{q_{\rm min}^2} - \frac{q_0^4}{q_{\rm max}^2} & , & \FDMq = \left(q_0/q\right)^2 \\
\end{array}\right.
\eeq
where all the $\varphi_c$ dependence is within $q_{\rm min}$ and the maximal momentum transfer is either a cutoff depending on the experimental setup or a maximum of $q_{\rm max} = 2 \mu_{\chi 1} v_{\rm max}$ ($v_{\rm max}$ is the maximal DM velocity). Thus, close to DM mass threshold, when $q_{\rm min} \approx q_{\rm max}$, the modulations dependence on $\FDMq$ vanishes, while for $q_{\rm min} < q_{\rm max}$, modulation for $\FDMq = \left(q_0/q\right)^2$ is enhanced with respect to the trivial DM FF.


\subsection{Neutrino-Nucleus Scattering Rate for Crystal Targets}
The differential neutrino-nucleon, neutral current cross-section is given by,
\begin{eqnarray}
\frac{\partial \sigma}{\partial q^2 \partial\varphi^{\prime\prime}_{q}} & = & \frac{G^2}{16\pi^2} \Big[ (G_V+G_A)^2 + (G_V-G_A)^2\left(1-\frac{q^2}{2ME_\nu}\right)^2 \nonumber \\
& & - (G_V^2-G_A^2)\frac{M q^2}{2ME_\nu^2} \Big] \,,
\end{eqnarray}
where $E_\nu$ is the energy of the incoming neutrino, $G$ is the Fermi constant and the angle $\varphi^{\prime\prime}_{q}$ now defines the rotation of $\mathbf{q}$ around the direction of the incoming neutrino ($z^{\prime\prime}$ in the double prime coordinate system now points in the direction of the propagation of the incoming neutrino). The parameters $G_V,G_A$ are the vector and axial contributions to the hadronic current, respectively. These can be parameterized as,
\begin{eqnarray}
G_V & \equiv & \left[g_V^p Z + g_V^n N \right] F_{\rm nuc}^V(q^2)\,, \nonumber \\
G_A & \equiv & \left[g_A^p (Z_+ - Z_-) + g_A^n (N_+ - N_-) \right] F_{\rm nuc}^A(q^2)  \,,
\end{eqnarray}
where $Z$ ($N$) is the number of protons (neutrons) and $Z_{\pm}$ ($N_{\pm}$) are the number of protons (neutrons) with spin plus or minus. For our purposes, the form factors  $F_{\rm nuc}^{V/A}(q^2)$ are just unity, since the momentum transfer is typically small. The appropriate values of the coefficients $g_V^p$, $g_V^n$, $g_A^p$, and $g_A^n$ can be found in \cite{Freedman:1977xn}.

The kinematically allowed range of momentum transfer is now,
\beq
q_{\rm min}(\Omega^\prime_q) \leq q \leq 2 E_\nu,
\eeq 
with the single prime coordinate system defined as before. Thus, the differential scattering rate for coherent neutrino-nucleus neutral current interactions is,
\beq
\frac{\partial R_\nu}{\partial q^2} = N_T \int d E_\nu \int d \varphi^{\prime\prime}_q \frac{\partial F}{\partial E_\nu} \frac{\partial \sigma}{\partial\varphi^{\prime\prime}_q \partial q^2} \Theta(q - q_{\rm min}(\Omega^\prime_q)),
\label{eq:Nu_Rate_1}
\eeq
where $\partial F/\partial E_\nu$ is the differential neutrino flux.

The solid angle $\Omega^\prime_q$ now depends on the same parameters as in the DM case, except that the dependence on DM velocity can be translated to a dependence on the value of the incoming neutrino energy and the direction of its propagation. For the case of Solar neutrinos, this just depends on the direction of the Sun. Taking the direction $\hat{x}^{\prime\prime}$ to be the Ecliptic North Pole and $-\hat{z}^{\prime\prime}$ to be the direction of the Sun, defines the direction from which the neutrino hits the detector, $\varphi_s$, (which is the angle of rotation of $\hat{z}^{\prime\prime}$ around $\hat{x}^{\prime\prime}$) and can be evaluated in the coordinate system of the target. Therefore, the $\mathbf{q}$ direction can be written as a function of the variables $\Omega^\prime_q = \Omega^\prime_q(q,E_\nu,\varphi_q^{\prime\prime},\varphi_s,\varphi_c)$. Thus, Eq.~(\ref{eq:Nu_Rate_1}) can be rewritten in the form of Eq.~(\ref{eq:dRnudlnEr}) with an explicit dependence on $\varphi_s$ and $\varphi_c$.


\bibliographystyle{ieeetr}
\bibliography{CCs_Proposal.bib}

\end{document}